# Mid-Infrared Radiative Emission from Bright Hot Plasmons in Graphene

Laura Kim[1,†], Seyoon Kim[1,3,†], Pankaj K. Jha[1], Victor W. Brar[1,2,3], Harry A. Atwater[1,2,*]

1. Thomas J. Watson of Applied Physics, California Institute of Technology, Pasadena, California 91125, USA
2. Kavli Nanoscience Institute, California Institute of Technology, Pasadena, California 91125, USA
3. Department of Physics, University of Wisconsin-Madison, Madison, WI 53711, USA

[*]haa@caltech.edu
[†]These authors contributed equally to this work.

## Abstract

The decay dynamics of excited carriers in graphene have attracted wide scientific attention, as the gapless Dirac electronic band structure opens up relaxation channels that are not allowed in conventional materials. We report Fermi-level-dependent mid-infrared emission in graphene originating from a previously unobserved decay channel: hot plasmons generated from optically excited carriers. The observed Fermi-level dependence rules out Planckian light emission mechanisms and is consistent with the calculated plasmon emission spectra in photoinverted graphene. Evidence for bright hot plasmon emission is further supported by Fermi-level-dependent and polarization-dependent resonant emission from graphene plasmonic nanoribbon arrays under pulsed laser excitation. Spontaneous plasmon emission is a bright emission process as our calculations for our experimental conditions indicate that the spectral flux of spontaneously generated plasmons is several orders of magnitude higher than blackbody emission at a temperature of several thousand Kelvin. In this work, it is shown that a large enhancement in radiation efficiency of graphene plasmons can be achieved by decorating graphene surface with gold nanodisks, which serve as out-coupling scatterers and promote localized plasmon excitation when they are resonant with the incoming excitation light. These observations set a framework for exploration of ultrafast and ultrabright mid-infrared emission processes and light sources.

Carrier relaxation in graphene is now understood to occur via several stages and decay channels. The promptly excited carriers with a non-Fermi-like distribution undergo carrier-carrier and carrier-plasmon scatterings on a 10-fs timescale, followed by Auger recombination and optical phonon emission. Excited carriers eventually reach a complete equilibrium with the lattice and environment through direct or disorder-assisted acoustic phonon emission processes, which occur on a picosecond timescale. These carrier relaxation processes in graphene upon optical pumping are depicted in Fig. 1a (refs 1-5). Previous studies have predicted and revealed the strong interplay of plasmons and particle/hole excitations in graphene, which plays a significant role in reducing the lifetime of photoexcited charge carrier[6-9]. Several theoretical studies have proposed that plasmon emission is another competing decay channel[10,11]. Doping graphene is not the only mechanism to generate plasmons, but plasmons can be generated as a result of decay of photo-excited carriers in graphene. Experimental evidence for optically generated non-equilibrium plasmons was provided by near-field microscopy measurements, where an increase in the Drude weight and the form of the resultant plasmon dispersion relations were consistent with graphene plasmons at an elevated carrier temperature[12,13].

In this paper, we report laser-pumped mid-infrared radiative emission from graphene, originating from bright hot plasmons emitted via decay of excited carriers. The observed Fermi-level dependence is consistent with the predicted plasmon emission spectra originating from photoinverted graphene achieved upon 100-fs pulsed excitation. This laser-pumped plasmon emission process can provide bright stimulated and spontaneous light emission sources. Our calculations suggest that there exist conditions of plasmon gain on a sub-100 fs timescale, during which stimulated plasmon emission dominates spontaneous plasmon. This suggests the intriguing future possibility of achieving coherent graphene plasmon amplification on this timescale. At times >100 fs, spontaneous plasmon emission contributes most to the overall mid-infrared light emission. As a result, the cumulative emission per laser excitation pulse is dominated by spontaneous plasmon emission. Spontaneous plasmon emission is a bright emission process as our calculations for our experimental conditions suggest that the spectral flux of spontaneously generated plasmons in graphene can be several orders of magnitude higher than blackbody emission at a temperature of several thousand Kelvin. A further experimental evidence for bright spontaneous plasmon emission is provided via Fermi-level-dependent and polarization-dependent emission measurements from periodic arrays of graphene nanoribbons, which are previously shown to support strong plasmonic resonances in mid-infrared[14-17] and allow efficient coupling to mid-infrared radiation[18-21].

Our experimental configuration is shown in Fig. 1b. The sample was illuminated with 100-fs pulses from a Ti:Sapphire laser operating at a wavelength of 850 nm while the Fermi level of graphene was externally controlled via electrostatic gating. The resulting emission spectra were collected using a Fourier transform infrared (FTIR) microscope, which acquired emission from a 50 μm×50 μm sample area (see Methods). Laser pulses arrive periodically at the sample approximately every 12 ns, while the moving mirror of the FTIR moves on a millisecond timescale. Thus, the repetition rate of the laser is sufficiently high that a large number of pulse-induced radiation events are integrated in each acquired spectrum.

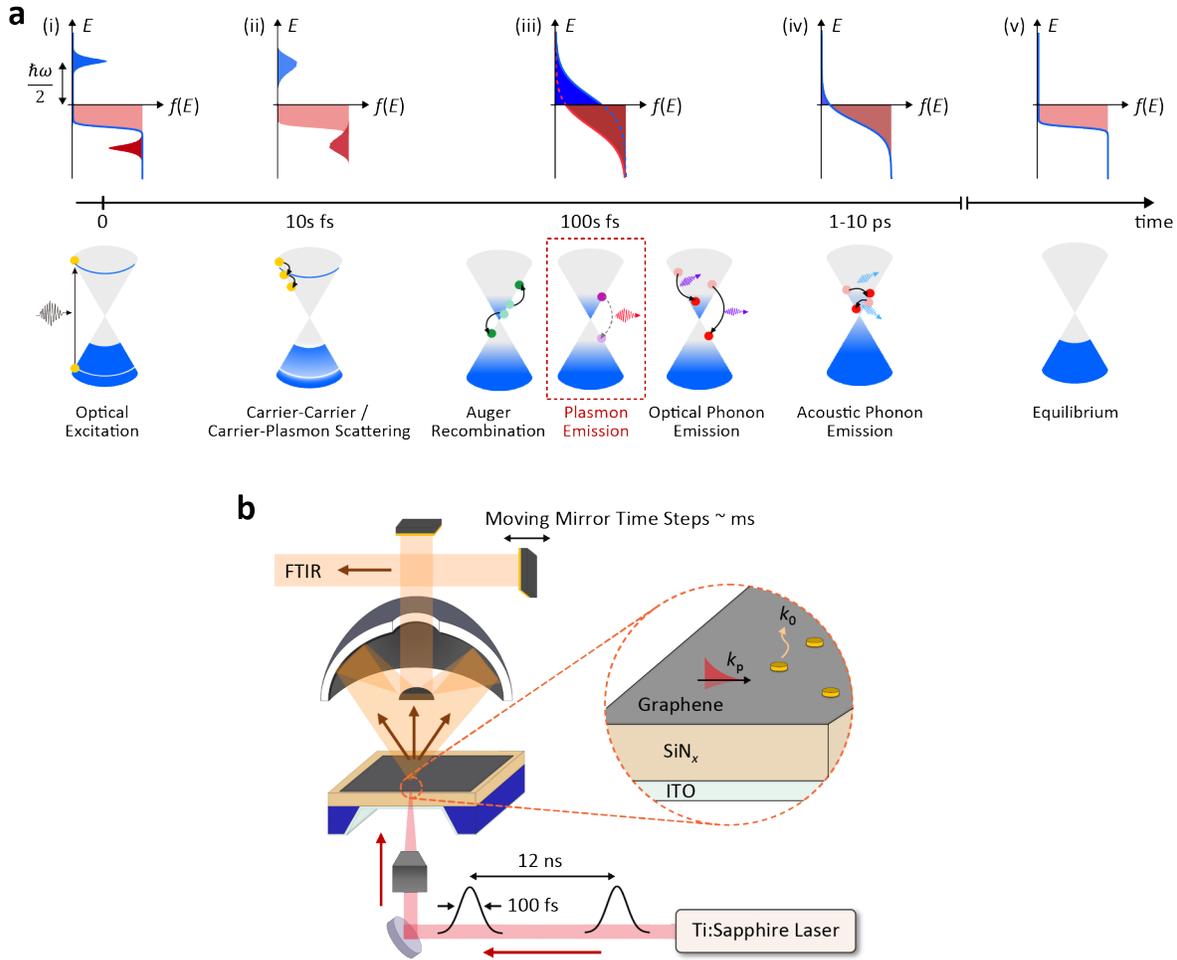

**Figure 1 | Carrier relaxation processes in graphene and experimental configuration.** (**a**) Carrier relaxation processes in graphene under ultrafast optical excitation: (i) Sharply peaked distribution of photoexcited carriers upon optical pumping. (ii) Carriers with a non-Fermi-like distribution undergoing carrier-carrier scattering on a 10-fs timescale. (iii) Carriers in a quasi-equilibrium state. (iv) Carriers that have been thermalized under interband processes, but are still hotter than the lattice. (v) Complete equilibrium between the carriers and the lattice. (**b**) Far-field infrared emission measurement setup.

Fermi-level-dependent emission spectra under pulsed laser excitation with a constant fluence of 1.12 J m$^{-2}$ are shown in solid color lines in Fig. 2a. Graphene Fermi levels denoted in the figures of this report are the gate-controlled Fermi levels of holes determined under equilibrium conditions. An increase in emission intensity with increasing graphene Fermi level was observed between 4.5 μm and 8 μm under pulsed laser excitation. Notably, when compared to the measured thermal emission spectrum at 95 °C (black dotted line in Fig. 2a), the emission spectra under pulsed laser excitation show gate-dependent excessive emission between 4.5 μm and 8 μm. The gate-dependent thermal emission spectra are shown in solid curves in Fig. 2b, and they were obtained by multiplying the Planckian emission spectrum of a blackbody by the measured absorptivity (or emissivity) of our device. Not only the spectral features observed under pulsed laser excitation are distinctively different from those seen in the thermal emission spectra, but also the gate-dependent trend is opposite. The gate-dependent thermal emission shows a decrease in emission intensity with an increasing graphene Fermi level between 8 μm and 12 μm, whereas the 100-fs-laser-induced emission shows an increase in emission intensity between 4.5 μm and 8 μm.

In our time-integrating measurement setup, all possible mid-infrared emission processes occurring under optical excitation are collected. Possible mid-infrared light emitting mechanisms include Planckian emission processes that describe direct photon emission whose spectral features are dictated by the temperatures of emitters, such as thermal emission due to substrate heating. Ultrafast photoluminescence, in which excited carriers with well-defined temperature emit with spectral features consistent with Planck's law, was previously shown be a light emitting mechanism in graphene under pulsed laser excitation[22]. In order to understand how the time evolution of carrier temperatures on ultrafast timescales may be manifested in an observed time-averaged emission spectrum, a phenomenological two-temperature (2T) model was adopted[22,23]. Time-dependent emission contributions from all three layers of the device were calculated and were added according to $\sum_{i=\text{graphene,ITO,SiN}_x} \xi_i(T_i(t)) \times S(\lambda, T_i(t))$, where $\xi$ is the temperature-dependent emissivity and $S$ is the blackbody spectral radiance given by Planck's law. The results were time-averaged and are shown as dotted curves in Fig. 2b. The gate dependence observed in the ultrafast Planckian emission between 4.5 μm and 8 μm is opposite to that seen in the measured spectra under pulsed optical excitation. Having control over Fermi levels of graphene via electrostatic gating generates qualitative gate-dependent trends and enables elimination of the Planckian radiation processes for the observed emission under pulsed laser excitation.

Plasmon emission is another light emitting mechanism as optically generated plasmons can scatter out in to the far-field in the form of mid-infrared light. To confirm photogeneration of graphene plasmons originating from inverted carriers in graphene, the observed light emission spectra under pulsed laser excitation are compared with the spectra taken under continuous wave (CW) laser excitation of the equivalent average laser power (Fig. 2c). Under CW laser excitation, the gate dependence occurs at longer wavelengths (>8 μm) similar to that seen in thermal emission under isothermal conditions, and the spectral shape is in good agreement with that of the measured thermal emission spectrum at 95 °C. Thus, we conclude that the major contribution of the observed emission under CW laser excitation is thermal emission. We note that the emission behavior seen under pulsed laser excitation was not observed with CW laser excitation. The observed emission behavior under pulsed laser excitation relies on the fact that carriers are excited with a sufficiently high carrier generation rate to achieve inversion. Considering a peak power of a 100-fs laser pulse, which is approximately $1.25 \times 10^5$ times higher than the average power, the carrier generation rate is expected to be approximately five orders of magnitude higher under pulsed laser excitation than that under CW laser excitation.

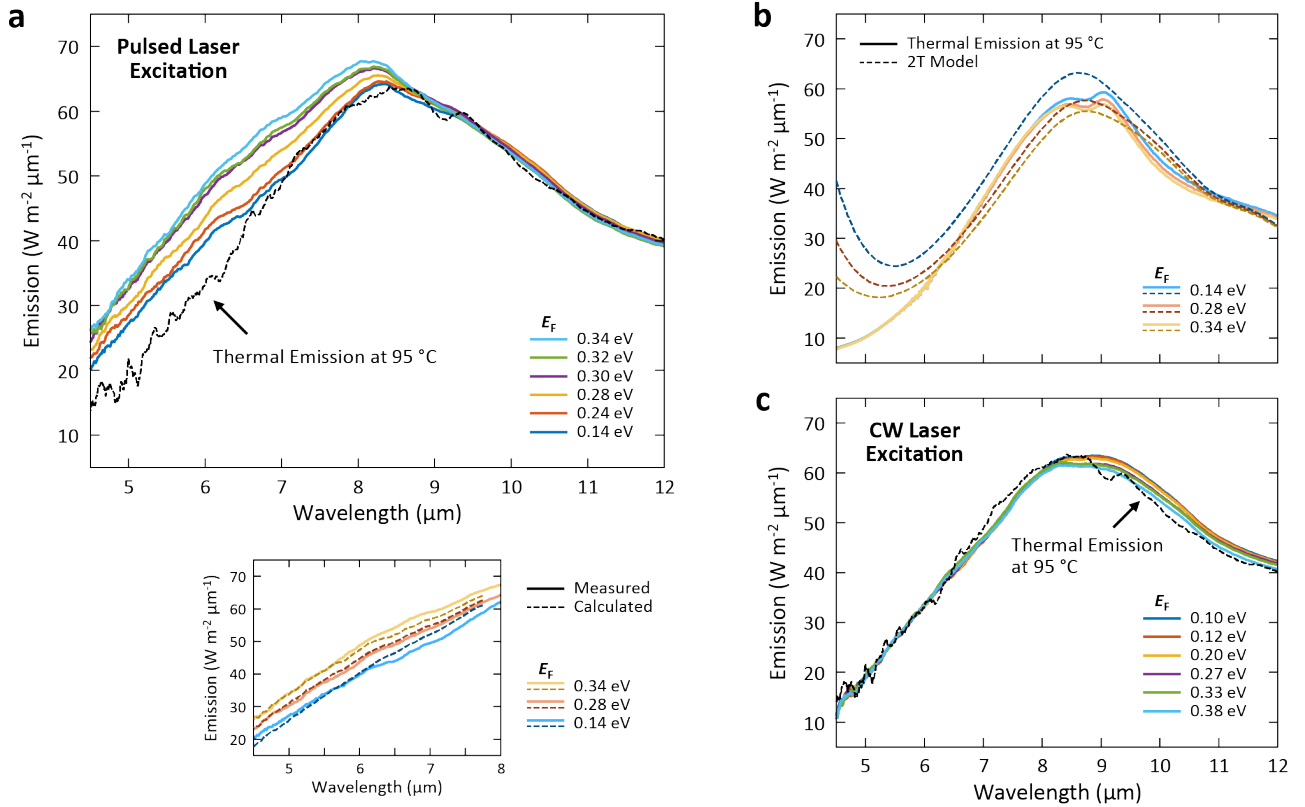

**Figure 2 | Graphene-Fermi-level-dependent emission spectra in a planar graphene.** (**a**) Measured Fermi-level-dependent emission spectra under pulsed laser excitation with a constant fluence of 1.12 J m$^{-2}$ (solid color lines), compared with the measured thermal emission spectrum at 95 °C (dotted black line). Comparison between the measured spectra and the calculated out-coupled spontaneous plasmon emission spectra is present in the bottom inset. (**b**) Fermi-level-dependent Planckian emission spectra under isothermal (solid color lines) and time-dependent 2T (dotted color lines) conditions. (**c**) Measured Fermi-level-dependent emission spectra under continuous wave laser excitation with a constant fluence of 1.12 J m$^{-2}$ (solid color lines), compared with the measured thermal emission spectrum at 95 °C (dotted black line).

Having known that the observed emission phenomena arise from optically generated graphene plasmons upon 100-fs laser pulses, the plasmon emission rates originating from photoinverted graphene and resultant plasmon emission spectra are calculated using Fermi's golden rule (FGR). As emission/absorption rates critically depend on the exactness of the plasmon dispersion relations[24], in order to accurately describe non-equilibrium plasmons, we employ a theoretical formulation to calculate exact complex-frequency plasmon dispersion with an arbitrary non-equilibrium carrier distribution[24,25]. The complex graphene plasmon dispersion relation, $\omega(k)=\omega_p(k)+i\gamma_p(k)$, where $\omega_p$ is the plasmon energy for a given $k$, and $\gamma_p$ is the plasmon decay rate[24,25], is solved exactly by setting the dielectric function, $\varepsilon(\omega,k)$, to zero. The dielectric function, $\varepsilon(\omega,k)$, can be expressed within the random phase approximation as Eq. (3).

$$\varepsilon(\omega, k) = 1 - \frac{e^2}{2\varepsilon_0 \varepsilon_{\text{eff}} k} \Pi_{\text{final}}(\omega_p + i\gamma_p, k) \tag{3}$$

where $\varepsilon_{\text{eff}} = \frac{1+\varepsilon_{\text{SiN}_x}}{2}$ is the average dielectric function of the air-SiN$_x$ interface, $\varepsilon_{\text{SiN}_x}$ is the complex dielectric function of SiN$_x$, and $\Pi_{\text{final}}(\omega,k)$ is the graphene dynamical polarizability. Because the carrier-carrier scattering is typically one to two orders of magnitude faster than the interband recombination processes[4,5], it is assumed that the system achieves a quasi-equilibrium state (equivalent to stage(iii) in Fig. 1a). In a quasi-equilibrium state, the carriers form a two-component plasma by having separate Fermi-Dirac distributions within their bands with their own chemical potentials, but the shared temperature. The polarizability for the two-component plasma system with finite temperatures is defined as the sum of the zero-temperature quasi-equilibrium polarizability and the correction terms that account for smearing of the Fermi edge as follows[24]:

$$\Pi_{\text{final}} = \Pi(\omega,k)|_{\text{quasi-eq}}^{T=0} + \int_0^\infty dE \left[ \frac{\partial \Pi|_{\mu=E}^{T=0}}{\partial E}(\delta f|_{\mu_c}^T) + \frac{\partial \Pi|_{\mu=E}^{T=0}}{\partial E}(\delta f|_{\mu_v}^T) \right] \tag{4}$$

where $\Pi(\omega,k)|_{\text{quasi-eq}}^{T=0}$ is the zero-temperature quasi-equilibrium polarizability, $\mu_c$ and $\mu_v$ are the chemical potentials for conduction and valence bands, respectively, $T$ is the shared temperature of the two-component plasma, and $\delta f = f(E)_\mu^T - f(E)_\mu^{T=0}$, where $f(E) = \frac{1}{e^{(E-\mu)/k_B T}+1}$. According to our analysis of contributions of stimulated and spontaneous plasmon emission processes per pulse as a function of time, the overall time-integrated emission per pulse is dominated by spontaneous plasmon emission. The spontaneous emission spectra are then calculated by scaling the plasmon emission rates by the plasmon density of states, $G=gD_p$, where $D_p = \frac{k(\omega)}{2\pi}\frac{dk(\omega)}{d\omega}$, assuming that the generated plasmons emit incoherently into all possible modes. To directly compare the calculated plasmon emission and the measured light emission, the out-coupling efficiencies of graphene plasmons were calculated based on the experimentally determined surface roughness of the sample. The radiation efficiency of plasmons in a planar graphene is found to be in the order of 10$^{-4}$. The out-coupled Fermi-level-dependent spontaneous plasmon emission spectra show good agreement with the measured emission under pulsed laser excitation as shown in the inset of Fig. 2a (dotted color lines). The increase in emission with increased hole doping of graphene can be intuitively understood as a result of enlarging the phase space for the excited carriers to relax by emitting plasmons.

We note that the gate dependence in the observed emission between 4.5 μm and 8 μm under pulsed laser excitation arises only from graphene. It has been previously reported that applying a sufficiently high gate voltage of order 1 V per 1 nm to ITO can yield a charge accumulation layer in ITO[26]. However, the charge accumulation in the ITO layer is negligible in our experimental conditions as the applied electric field is an order of magnitude smaller than that required to induce an ITO accumulation layer[26]. As an independent check, a control experiment was performed with a graphene-less control sample, consisting of a 1 μm-thick SiN$_x$ sandwiched between 50 nm-thick ITO layers, which served as top and bottom gates. When illuminated with pulsed laser excitation with a fluence of 1.12 J m$^{-2}$, negligible gate dependence was observed.

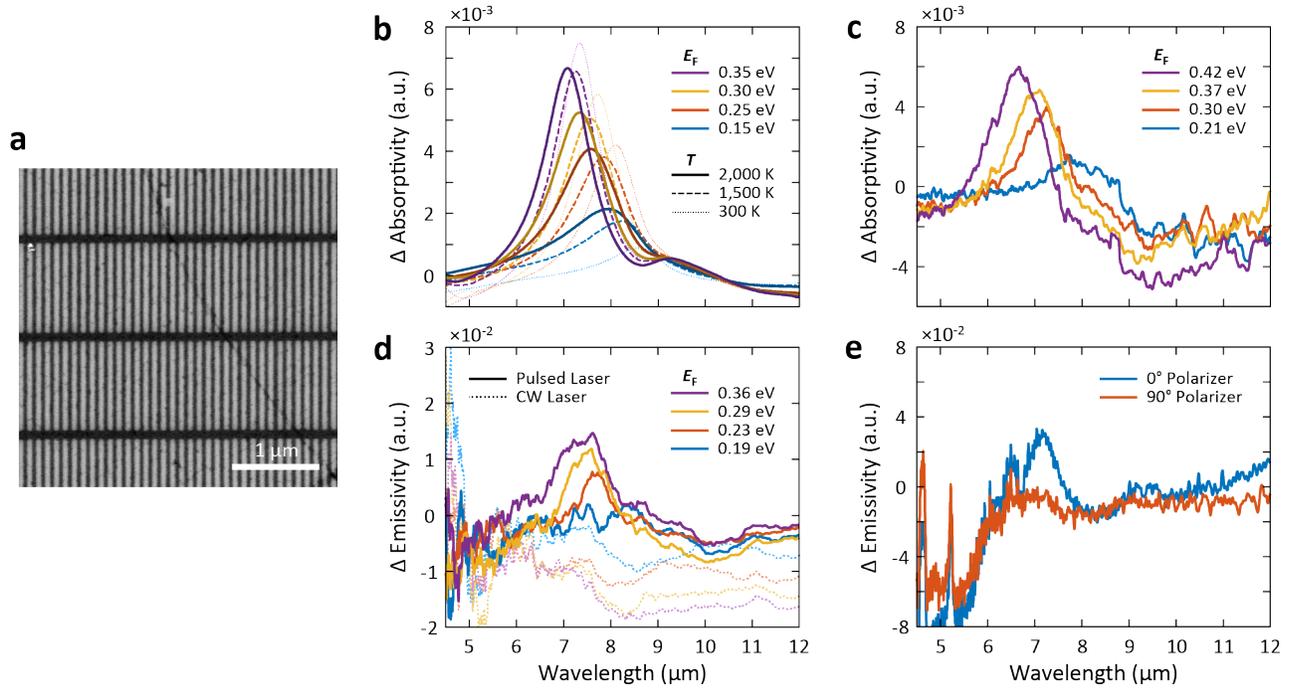

**Figure 3 | Graphene-Fermi-level-dependent emission spectra from 35 nm graphene nanoribbon arrays.** (**a**) SEM image of 35 nm graphene nanoribbon arrays. (**b**) Calculated graphene-Fermi-level-dependent plasmonic resonant changes in absorptivity from periodic arrays of 35 nm graphene nanoribbons with varying carrier temperatures in graphene. (**c**) Measured graphene-Fermi-level-dependent changes in absorptivity from periodic arrays of 35 nm graphene nanoribbons at room temperature. (**d**) Measured gate-dependent changes in emissivity from periodic arrays of 35 nm graphene nanoribbons under pulsed (solid lines) and CW (dotted lines) laser excitations. (**e**) Measured polarization-dependent changes in emissivity under pulsed laser excitation for a gate-controlled graphene Fermi level of 0.35 eV.

To further demonstrate that plasmons are the source of the detected gate-dependent light emission, we also performed emission measurements on periodic arrays of graphene nanoribbons, which are well-known to support localized graphene plasmonic resonances[14-17]. Graphene nanoribbons, whose width and pitch are 35 nm and 100 nm, respectively, were fabricated on equivalent substrates (i.e., 1 µm-thick SiN$_x$ and 50 nm-thick ITO), as shown in Fig. 3a. As standing-wave-like localized plasmonic modes are excited across the width of the graphene nanoribbons, the resonant conditions are determined by the width of the nanoribbons and the graphene Fermi level through the relationship $\omega_p \propto \sqrt{E_F}/\sqrt{W}$, where W is the graphene nanoribbon width. For a fixed nanoribbon width of 35 nm, the plasmonic resonant behavior is demonstrated by the measured Fermi-level-dependent changes in absorptivity (i.e., absorptivity measured when graphene is at CNP is subtracted from absorptivity when graphene is gated) as shown in Fig. 3c. The intensity and frequency of these plasmonic modes blue shift with increasing graphene Fermi level. The electromagnetic calculations confirm the plasmonic origin of the observed resonant features in the absorptivity measurements (Fig. 3b). As shown in the calculated results, the resonant features slightly blueshift with increasing carrier temperature as the Drude weight of graphene, which represents the oscillator strength of free carriers, is also a function of carrier temperature[12,27]. The gate-dependent emission spectra from the 35 nm graphene nanoribbon arrays under pulsed laser excitation are shown in solid lines in Fig. 3d. They display apparent resonant features that show strong correlation with the plasmonic absorption features. We note that no resonant peaks were observed under CW laser excitation as shown in the dotted lines in Fig. 3d. The resonant features observed in the emission spectra under pulsed laser excitation, but not under CW laser excitation, indicate that only under pulsed excitation are the optically generated plasmons out-coupled at the graphene nanoribbon resonant frequency. As discussed previously, under CW laser excitation, thermal radiation gives the largest contribution to the measured emission. The observed phenomena under ultrafast optical excitation are distinctly different from the thermal emission effects from similar graphene nanoribbon structures shown in ref. 19. Additional experimental evidence for plasmon emission is provided via the polarization-dependent measurements shown in Fig. 3e. The feature that appeared under pulsed laser excitation near 7 μm is strongly polarized, as is expected for laterally confined graphene plasmonic resonant modes. This feature vanishes quickly as we rotate the polarization of the probing radiation from 90° to 0° relative to the graphene nanoribbon axis.

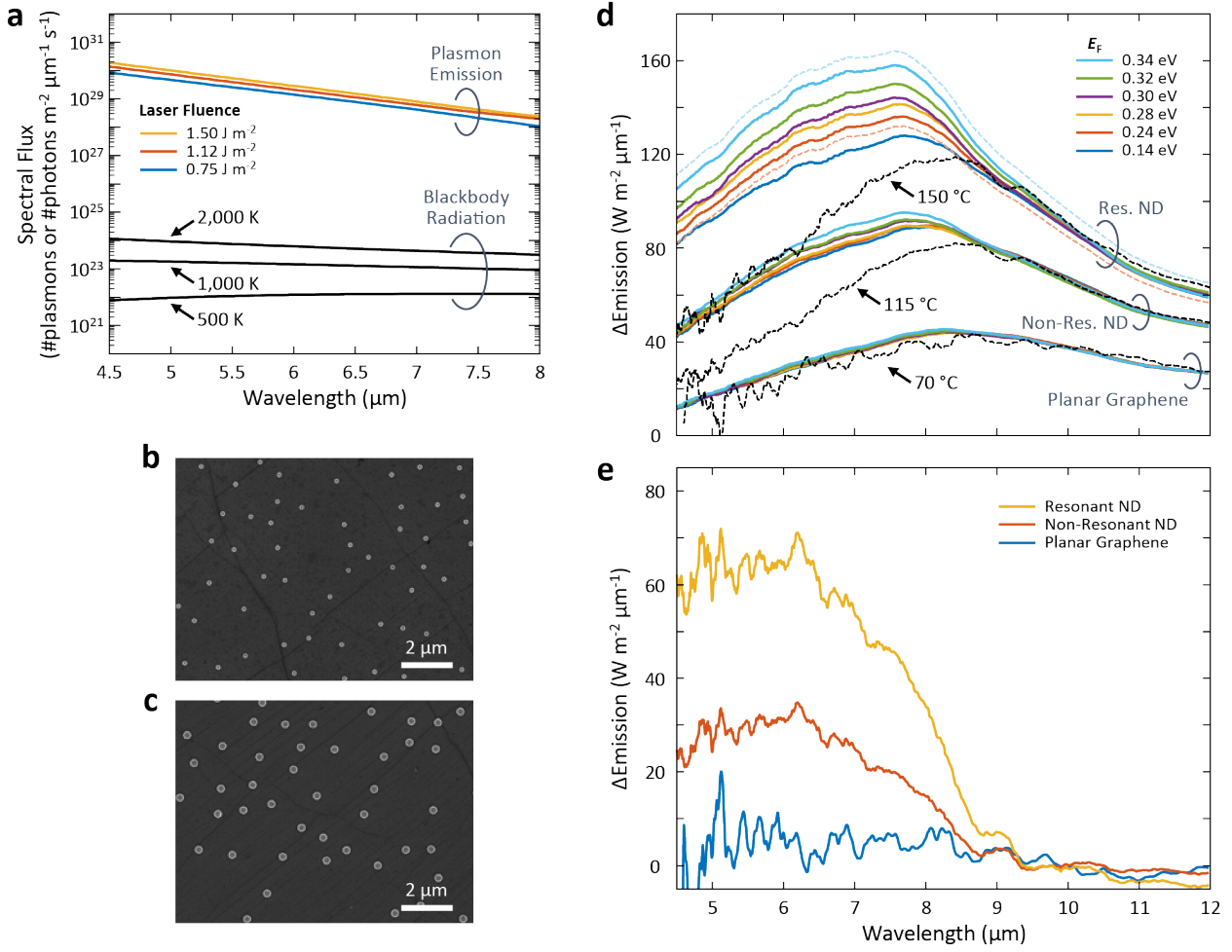

**Figure 4 | Bright spontaneous plasmon emission and enhancing radiation efficiency of plasmons.** (a) Spectral flux of cumulative spontaneous plasmon emission assuming unity out-coupling efficiency at various laser fluences for a given graphene Fermi level of 0.34 eV compared with the spectral flux of blackbody radiation at 500 K, 1,000 K, and 2,000 K. **(b), (c)** SEM images of randomly distributed gold nanodisks on graphene that are resonant (top) and non-resonant (bottom) with the incoming laser excitation, respectively. **(d)** Fermi-level-dependent emission spectra from planar and ND-decorated graphene samples under pulsed laser excitation with a constant laser fluence of 0.75 J m$^{-2}$ (color solid lines). In the resonant ND data, the dotted lines for 0.24 eV and 0.34 eV correspond to the original measurements, and the solid lines are fitted to compensate the offset, which appeared due to experimental imperfections. Measured thermal emission spectra from the device for given temperatures of 70 °C, 115 °C, and 150 °C (black dotted lines). **(e)** Emission contributions that deviate from the corresponding thermal emission background under pulsed laser excitation with a constant laser fluence of 0.75 J m$^{-2}$ from planar graphene and ND-decorated graphene samples at a given gate-controlled graphene Fermi level of 0.34 eV.

Spontaneous plasmon emission originating from photoinverted graphene is a bright emission process. As shown in Fig. 4a, the calculated cumulative spectral flux of spontaneously emitted plasmons per pulse is several orders of magnitude higher than that of photons emitted by a blackbody at several representative temperatures of 500 K, 1,000 K, and 2,000 K. The plasmon emission process can provide a platform for achieving ultrabright mid-infrared spontaneous light sources.

Here we demonstrate that the far-field radiation efficiency of spontaneously generated graphene plasmons can be greatly enhanced by placing engineered nanostructures on graphene surface. Gold nanodisks (NDs) that are either resonant or non-resonant with the laser excitation wavelength were fabricated on planar graphene. As shown in Figs. 4b and c, they were randomly distributed over graphene surface with the surface coverages of approximately 1% and 2.8% for the resonant and non-resonant NDs, respectively. Figure 4d shows the measured emission spectra under pulsed laser excitation from planar and ND-decorated graphene samples. At long wavelengths (>8 μm), the observed emission spectra match well with the measured thermal emission profiles, with the ND-decorated samples reaching apparent higher temperatures than the planar graphene sample. This suggests that the gold NDs act as local heating source due to their absorption at 850 nm. However, in addition to those heating effects, the resonant and non-resonant NDs cause a large gate-dependent deviation from the thermal emission profiles between 4.5 μm and 8 μm. Figure 4e shows the spectra under pulsed laser excitation for a given gate-controlled graphene Fermi level of 0.34 eV after the corresponding thermal emission background is subtracted.

The samples with resonant and non-resonant NDs display seven and three times larger emission relative to their corresponding thermal backgrounds, respectively, than the planar graphene sample. While localized heating near the NDs could lead to a non-uniform temperature across the sample, the large gate dependence at shorter wavelengths (<8 µm) in the presence of NDs suggests that their dominant effect is to more efficiently generate and/or out-couple hot-carrier generated graphene plasmons (i.e, a given change in plasmon emission with changing $E_F$ × higher out-coupling efficiency = larger out-coupled gate dependence). Furthermore, the large gate-dependent emission seen between 4.5 µm and 8 µm is bigger with resonant NDs than with non-resonant NDs. The strong near-field enhancement created by resonant NDs impacts both population inversion and plasmon emission rates. The resonant NDs enhance absorption in graphene locally, creating more excited carriers in the vicinity of the NDs, locally enhancing carrier inversion[20]. It also has been previously reported that resonantly excited NDs can inject carriers into graphene[28,29], effectively changing the doping level of graphene. While this process is difficult to quantitatively assess under our conditions, any hot carrier doping process is likely to contribute to increased plasmon emission. Fermi level pinning for graphene in contact with a metal can locally change the effective chemical potential[30], which could result in more plasmons emitted near the NDs. The localized plasmon excitations scatter out more efficiently as they are excited near a ND.

In conclusion, our experimental results suggest that quasi-equilibrium 'hot' carrier distributions in graphene upon ultrafast optical excitation support bright mid-infrared plasmonic excitation. The gate dependence provides the strongest evidence for plasmon emission as it rules out all of the other infrared light emitting mechanisms and suggests that the origin of the observed emission phenomena is optically generated plasmons in photoinverted graphene. For plasmon emission being an interband process, having control over graphene Fermi level via electrostatic gating enhances observation of plasmon-coupled radiation, as hole-doping of graphene enlarges phase space for plasmon emission and also raises the (Pauli-blocking) barrier for plasmon absorption. The novel claim of bright spontaneous plasmon emission is further supported by gate-dependent and polarization-dependent resonant emission features observed from 35 nm graphene plasmonic nanoribbon arrays. Furthermore, we show that the far-field radiation efficiency can be greatly enhanced by engineering plasmon out-coupling structures on graphene.

In this work, the density of gold NDs was deliberately limited in order to ensure minimum changes in thermal emission characteristics. As increasing a density of the incorporated nanostructures will enhance radiation efficiency, lossless dielectric resonators can be explored as well as different substrates which can provide better heat sink. Ideally, an array of nanostructures separated by a distance that is comparable to the propagation length of plasmons will be able to efficiently out-couple plasmons before being damped. These analyses suggest that there are no inherent limitations of achieving a ultrabright, ultrafast plasmon-assisted light emission. These findings open a new path for exploration of mid-infrared stimulated and spontaneous emission processes and ultrafast and ultrabright light sources. In addition, optically excited plasmons could also be exploited the near field for applications that require intense localized mid-infrared intensity on the nanometer length scale.

## Methods

### Devices

The sample consists of a CVD-grown monolayer graphene layer on a 1 μm-thick $SiN_x$ and a 50 nm-thick ITO, which serves as a backgate. A 50 nm-thick ITO film was deposited on the bottom side of a 1 μm-thick $SiN_x$ membrane (Norcada, NX10500F) by RF sputtering with the flow rate of $Ar+O_2$ of 0.4 sccm at a pressure of 3 mTorr at a power of 48 W. CVD-grown monolayer graphene was transferred onto the top side of the $SiN_x$ membrane. The gold nanodisks were fabricated by patterning onto a PMMA resist on top of the graphene layer by 100 keV electron beam lithography and evaporating a 2 nm Ti followed by a 80 nm Au. The gold nanodisks that are resonant and non-resonant with the incoming laser light at the wavelength of 850 nm were fabricated with diameters of 170 nm and 285 nm, respectively. The nanodisks are randomly distributed in order to prohibit in-plane resonacnes originating from perfect periodicity. Over a 150 μm×150 μm area, 10,000 nanodisks were fabricated, covering approximately 1% and 2.8% of the graphene surface for the resonant and non-resonant nanodisks, respectively. In addition, the nanodisks are separated by at least 800 nm in center-to-center distance to suppress plasmonic interactions between the adjacent gold nanodisks. Therefore, each gold ND can be considered as an isolated single gold ND. Graphene nanoribbons with a width of 35 nm were patterned on a graphene surface using 100 keV electron beam lithography, followed by reactive ion etching with $O_2$.

### FTIR Emission Measurements

The sample was illuminated with sub-100 fs pulses from a Ti:Sapphire laser operating at 850 nm with a repetition rate of 80 MHz. The laser was focused onto the graphene surface from the backside. The emitted infrared light was collected with a 15× Cassegrain objective of a Fourier transform infrared (FTIR) microscope, and was sent to a Fourier-transform infrared (FTIR) spectroscopy before being focused on a liquid nitrogen-cooled HgCdTe detector. The spot size of the laser was large enough to ensure uniform illumination over the collection area of 50 μm×50 μm enclosed by an aperture. For thermal emission measurements, the device was placed on a temperature-controlled stage consisting of a 100 mm-thick sapphire on a 2 mm-thick copper on a heated silver block that can vary in temperature. The stage temperature was monitored via a thermocouple mounted in the silver block. All of the emission measurements were done under dry air purge. A black soot sample was used as an emissivity reference, and the collected emission spectra were calibrated assuming unity emissivity for the black soot reference at all wavelengths. For polarization-dependent measurements, a wire grid polarizer was placed in the collimated beam path.

**Acknowledgements**

This work was supported by U.S. Department of Energy (DOE) Office of Science Grant DE-FG02-07ER46405. S.K. acknowledges support by a Samsung Scholarship. P.K.J. also acknowledges support by the Boeing Company.

**Author contributions**

L.K., V.W.B. and H.A.A. conceived the ideas. L.K. performed spectroscopy experiments; L.K. performed inversion and gain calculations as well as emissivity calculations. L.K and S.K. fabricated the sample. L.K. and S.K. performed data analysis. P.K.J. contributed to the discussion of the ratio of stimulated to spontaneous emission rates calculations. All authors cowrote the paper, and H.A.A. supervised the project.